\PassOptionsToPackage{table,dvipsnames}{xcolor}
\documentclass[sigconf]{acmart}
\renewcommand\footnotetextcopyrightpermission[1]{} 
\usepackage[inline]{enumitem}

\AtBeginDocument{
  }
\usepackage{subcaption} 
\usepackage{array}
\usepackage[capitalise, nameinlink, noabbrev]{cleveref}
\usepackage{graphicx}   
\usepackage{float}
\usepackage{listings}
\lstdefinestyle{gram}{
  basicstyle=\ttfamily\fontsize{5.5}{5.6}\selectfont,
  columns=fullflexible,
  keepspaces=true,
  showstringspaces=false,
  breaklines=true,
  numbers=none,
  upquote=true,
  aboveskip=0pt,
  belowskip=0pt,
  moredelim=**[is][\color{red}]{@}{@},
}

\usepackage{tikz}
\usetikzlibrary{positioning,arrows.meta}
\usetikzlibrary{shapes,arrows}

\usepackage[most]{tcolorbox}
\usepackage{xcolor}
\usepackage{amsmath}
\usepackage{listings}
\usepackage{dot2texi}

\usepackage{amssymb} 
\usepackage{xcolor}
\usepackage{tabularx}  
\usepackage{caption}
\usepackage{booktabs}

\usepackage[textsize=tiny]{todonotes}

\usepackage{cleveref}
\usepackage{pgfplots}
\pgfplotsset{compat=1.18}

\usepackage{contour}
\usepackage[normalem]{ulem}

\contourlength{0.8pt}
\newcommand{\ul}[1]{
  \uline{\phantom{#1}}
  \llap{\contour{white}{#1}}
}

\newif\iflong
\longfalse 

\newcommand{\conclusion}[1]{
  \begin{center}
    \fbox{\begin{minipage}{0.95\columnwidth}
      \centering
      \emph{#1}
    \end{minipage}}
  \end{center}
}

\usepackage{wrapfig}
\usepackage{booktabs}   
\usepackage{makecell}
\usepackage{multirow}
\usepackage{algorithm}         
\usepackage[noend]{algpseudocode} 

\usepackage{xspace}
\newcommand{\AUTOSPEC}{\textsc{AutoSpec}\xspace}

\lstdefinelanguage{json}{
    basicstyle=\ttfamily\small,
    numbers=left,
    numberstyle=\tiny\color{gray},
    stepnumber=1,
    numbersep=5pt,
    showstringspaces=false,
    breaklines=true,
    frame=single,
    literate=
     *{0}{{{\color{numb}0}}}{1}
      {1}{{{\color{numb}1}}}{1}
      {2}{{{\color{numb}2}}}{1}
      {3}{{{\color{numb}3}}}{1}
      {4}{{{\color{numb}4}}}{1}
      {5}{{{\color{numb}5}}}{1}
      {6}{{{\color{numb}6}}}{1}
      {7}{{{\color{numb}7}}}{1}
      {8}{{{\color{numb}8}}}{1}
      {9}{{{\color{numb}9}}}{1}
}

\lstdefinelanguage{json}{
  basicstyle=\ttfamily,
  numbers=none,
  literate=
   *{0}{{{\color{black}0}}}{1}
    {1}{{{\color{black}1}}}{1}
    {2}{{{\color{black}2}}}{1}
    {3}{{{\color{black}3}}}{1}
    {4}{{{\color{black}4}}}{1}
    {5}{{{\color{black}5}}}{1}
    {6}{{{\color{black}6}}}{1}
    {7}{{{\color{black}7}}}{1}
    {8}{{{\color{black}8}}}{1}
    {9}{{{\color{black}9}}}{1},
}

\usepackage{pgf-umlsd}

\definecolor{clientblue}{RGB}{0,0,180}
\definecolor{serverred}{RGB}{180,0,0}

\newcommand{\diagramfontsize}{\scriptsize}

\lstdefinestyle{diagramlisting}{
  basicstyle=\ttfamily\diagramfontsize,
  frame=single,
  breaklines=true,
  columns=fullflexible,
  keepspaces=true,
  framesep=2pt,
  xleftmargin=1pt,
  framexleftmargin=1pt
}

\newcommand{\client}[1]{\textcolor{clientblue}{#1}}
\newcommand{\server}[1]{\textcolor{serverred}{#1}}
\newcommand{\exchange}[2]{
  \begin{call}{Client}{\client{#1}}{Server}{\server{#2}}\end{call}
}

\begin{document}

\title{Closing the Loop: Execution-Guided Synthesis of Formal Specifications from Natural Language}

\title{Synthesizing Precise Specifications from Natural Language for~Effective~Protocol~Fuzzing}
\title{Synthesizing Precise Protocol Specs from~Natural~Language for~Effective~Test Generation}

\author{Kuangxiangzi Liu $\cdot$ Dhiman Chakraborty}
\affiliation{
  \institution{Volkswagen AG}
  \city{Wolfsburg}
  \country{Germany}
}
\email{<first-name>.<last-name>@volkswagen.de}

\author{Alexander Liggesmeyer $\cdot$ Andreas Zeller}
\affiliation{
  \institution{CISPA Helmholtz Center for Information Security}
  \city{Saarbr\"ucken}
  \country{Germany}
}
\email{<first-name>.<last-name>@cispa.de}

\renewcommand{\shortauthors}{Trovato et al.}

\begin{abstract}

Safety- and security-critical systems have to be thoroughly tested against their specifications.
The state of practice is to have \emph{natural language} specifications, from which test cases are derived manually---a process that is slow, error-prone, and difficult to scale.
\emph{Formal} specifications, on the other hand, are well-suited for automated test generation, but are tedious to write and maintain.
In this work, we propose a two-stage pipeline that uses large language models (LLMs) to bridge the gap:
First, we extract \emph{protocol elements} from natural-language specifications; second, leveraging a protocol implementation, we synthesize and refine a formal \emph{protocol specification} from these elements, which we can then use to massively test further implementations.

We see this two-stage approach to be superior to end-to-end LLM-based test generation, as 
\begin{enumerate*}
\item it produces an \emph{inspectable specification} that preserves traceability to the original text;

\item the generation of actual test cases \emph{no longer requires an LLM};
\item the resulting formal specs are \emph{human-readable,} and can be reviewed, version-controlled, and incrementally refined; and
\item over time, we can build a \emph{corpus} of natural-language-to-formal-specification mappings that can be used to further train and refine LLMs for more automatic translations.
\end{enumerate*}

Our prototype, \AUTOSPEC, successfully demonstrated the feasibility of our approach on five widely used \emph{internet protocols} (SMTP, POP3, IMAP, FTP, and ManageSieve) by applying its methods on their \emph{RFC specifications} written in natural-language, and the recent \emph{I/O grammar} formalism for protocol specification and fuzzing.
In its evaluation, 
\AUTOSPEC recovers on average 92.8\% of client and 80.2\% of server message types, and achieves 81.5\% message acceptance across diverse, real-world systems---a first step towards automatic formalization of natural-language specifications for comprehensive test generation.

\end{abstract}



\maketitle

\section{Introduction}

At Volkswagen, continuous delivery depends on automated testing.
This is especially true in safety- and security-critical domains, where every revision must ship with confidence.
The state of the practice is that testing requirements are documented in natural language; engineers then have to turn these texts into executable checks by hand.
This process is slow, error-prone, and difficult to scale, creating inconsistencies and gaps in test coverage~\cite{RENLP}.
Recent developments in large language models suggest that it may be possible to automate this task and generate tests directly from natural language.
However, our experience asking LLMs to generate test cases reveals well-known weaknesses: hallucinated inputs, missed corner cases, poor traceability between requirements and tests, and most critically, a lack of systematic validation of the output---problems observed in multiple studies~\cite{llmHall,bender-koller-2020-climbing,llmhallu2,hallu3}.
These issues undermine assurance goals and make LLM-based testing difficult to trust in safety-critical settings~\cite{seAI}.

In this paper, we explore a different path.
Rather than asking an LLM to generate tests directly, we give it a narrower but more reliable task: translating \emph{natural-language requirements into formal, executable specifications,} an approach that has been proven successful in prior work to derive temporal logic specifications~\cite{nl2spec,Lang2LTL}.
Our approach is more ambitious, though: We target \emph{protocol specifications} that capture the \emph{entirety of interactions} between system components, including syntax, state, and formal constraints.
From such a specification, a test generator can deterministically produce interactions and oracles that systematically explore the input and output space of the component~\cite{iogrammars}.
\newlength{\tripleht}
\setlength{\tripleht}{4.8cm}

\begin{figure*}[!t]
  \centering
  \noindent\makebox[\textwidth][c]{
    
    \begin{subfigure}[b]{0.32\textwidth}
      \centering
      \begin{minipage}[b][\tripleht][t]{\linewidth}
        \includegraphics[width=\linewidth,height=\tripleht,keepaspectratio]{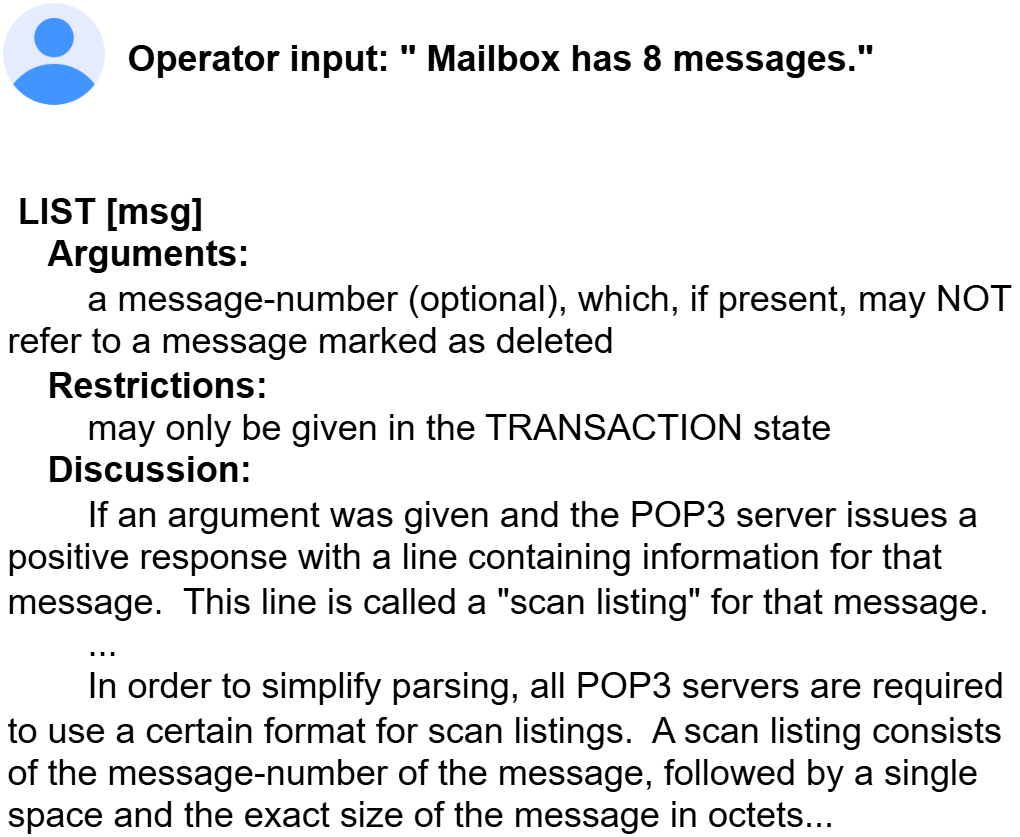}
      \end{minipage}
      \subcaption{POP3 RFC (excerpt)}
      \label{fig:triple:a}
    \end{subfigure}\hfill
    \hfill $\triangleright$ \hfill
    
    \begin{subfigure}[b]{0.32\textwidth}
      \centering
      \begin{minipage}[b][\tripleht][t]{\linewidth}
        \includegraphics[width=\linewidth,height=\tripleht,keepaspectratio]{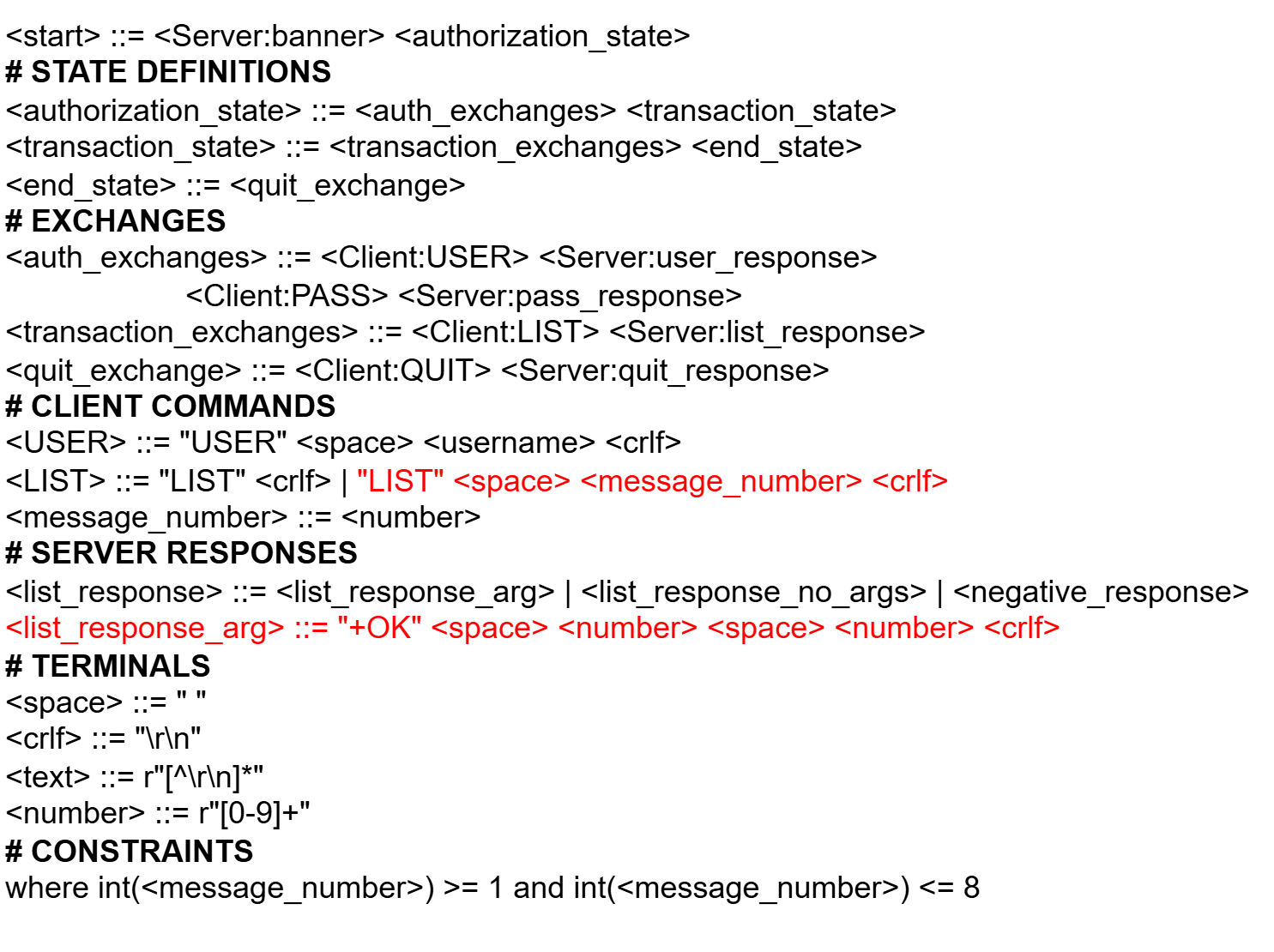}
      \end{minipage}
      \subcaption{POP3 I/O Grammar (excerpt)}
      \label{fig:triple:b}
    \end{subfigure}\hfill
    \hfill $\triangleright$ \hfill
    
    \begin{subfigure}[b]{0.32\textwidth}
      \centering
      \begin{minipage}[b][\tripleht][t]{\linewidth}
        \resizebox{\linewidth}{!}{
         \begingroup
              \renewcommand{\diagramfontsize}{\normalsize} 
              \begin{sequencediagram}\ttfamily\diagramfontsize
                \newthread{Client}{\client{\textnormal{\diagramfontsize Client}}}{}
                \newinst[5.2cm]{Server}{\server{\textnormal{\diagramfontsize Server}}}
                \exchange{(connect)}{+OK Dovecot ready.\textbackslash r\textbackslash n}
                \exchange{USER debug@localdomain.test\textbackslash r\textbackslash n}{+OK\textbackslash r\textbackslash n}
                \exchange{PASS NEWpass123\textbackslash r\textbackslash n}{+OK Logged in.\textbackslash r\textbackslash n}
                \exchange{LIST 8\textbackslash r\textbackslash n}{+OK 8 383\textbackslash r\textbackslash n}
                \exchange{QUIT\textbackslash r\textbackslash n}{+OK Logging out.\textbackslash r\textbackslash n}
              \end{sequencediagram}
            \endgroup
        }
      \end{minipage}
      \subcaption{Generated test trace}
      \label{fig:trace}
    \end{subfigure}
  }
  \caption[POP3 example]{From RFC text to I/O Grammar to test trace. We convert the natural-language POP3 protocol specification (a) into a formal I/O grammar (b). Instantiating this protocol specification yields concrete interactions as test cases (c).}
  \label{fig:triple}
\end{figure*}

As an example, let us consider the well-known POP3 protocol for retrieving email from a server.
As an internet standard, POP3 is well documented in an RFC document.
\Cref{fig:triple}(a) shows a snippet from RFC 1939~\cite{RFC1939}, which defines the \texttt{LIST} command to retrieve the headers of an email message.
The RFC specifies the syntax of the command, its expected behavior, and the possible responses---all in precise language, yet amenable to automated processing.

\Cref{fig:triple}(b) \emph{formalizes} this document into a so-called \emph{I/O grammar}~\cite{iogrammars}.
This grammar captures the output of the server (prefixed with \texttt{Server:}) as well as the output of the client (prefixed with \texttt{Client:}).
The exchange \texttt{<transaction\_exchanges>} (in the ``EXCHANGES'' section), for instance, specifies that after the POP3 client issues a \texttt{LIST} command (\texttt{<Client:LIST>}), the server responds with an adequate response (\texttt{<Server:list\_response>}).
Their syntax is defined further below in the grammar.

The grammar also captures the \emph{stateful} nature of the protocol: the \texttt{LIST} command can only be issued after authorization has taken place.
Furthermore, I/O grammars allow specifying \emph{semantic constraints} over message elements; the last line, for instance, specifies that the message number must be between 1~and~8---a property that cannot be expressed in a context-free grammar alone.
In protocols, satisfying such constraints is crucial for synthesizing correct inputs; and while the syntax of messages is often specified formally using grammars, their semantic constraints are not.

From the \emph{formal} specification in \Cref{fig:triple}(b), a fuzzer can  synthesize concrete \emph{test cases} that explore the interaction space of the protocol.
\Cref{fig:triple}(c) shows an example of such a generated test, which captures the sequence of messages exchanged between the client and server during a typical POP3 session.
Each such test can check a client or a server against the protocol specification, ensuring that it behaves correctly in all specified states and transitions, making the specification in \Cref{fig:triple}(b) very valuable.

Compared to the alternative of having an LLM generate tests directly from a natural-language specification, going via a formal specification has several advantages for us:
\begin{enumerate*}[label=(\arabic*)]
\item The specification is an \emph{inspectable artifact} that preserves sentence-level traceability to the original text~\cite{DBLP:conf/icse/Cleland-HuangGHMZ14};
\item All subsequent testing steps are reproducible and free of LLM nondeterminism, hallucinations, or cost;
\item Verification shifts from opaque model outputs to human-readable specs, which can be reviewed, version-controlled, and incrementally refined~\cite{TLA}; and
\item Over time, the collected specifications form a \emph{corpus} of natural-language-to-formal-specification mappings that can be used to further train and refine LLMs, and thus \emph{bootstrap} more automatic translations.
\end{enumerate*}

\begin{figure}[t] 
  \centering
  \includegraphics[width=1\linewidth]{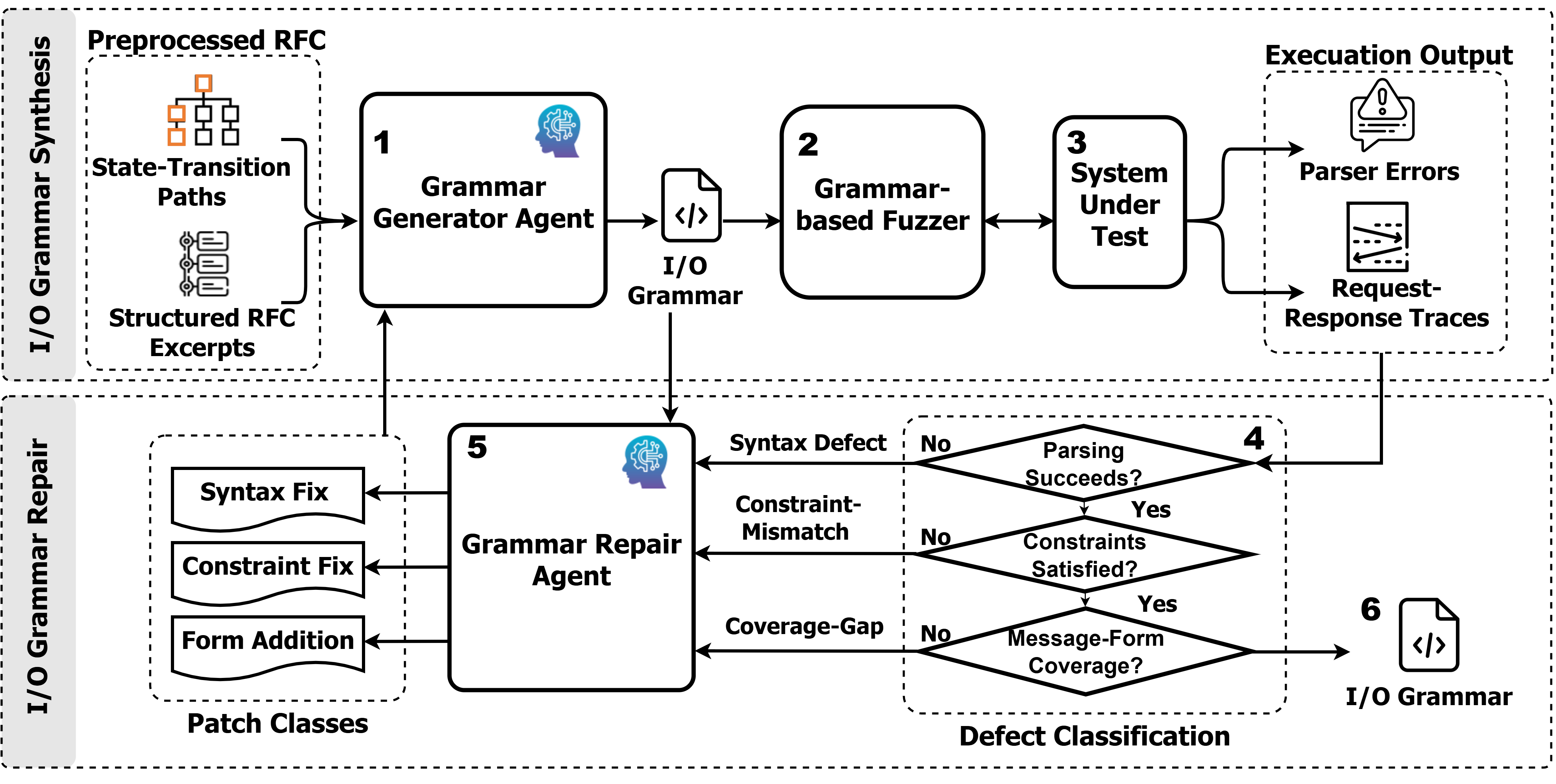}
  \caption[How \AUTOSPEC works.]{How \AUTOSPEC works. From protocol elements automatically extracted from an RFC, an LLM (1) produces an I/O grammar as formal protocol specification. A test generator (2) can then use this I/O grammar to test an actual implementation (3). If errors occur (4), a grammar repair agent (5) suggests specification fixes to the LLM, initiating a repair and refinement cycle. The result is a high-quality I/O grammar that can be used for comprehensive testing.}
  \label{fig:pipeline}

\end{figure}

How do we get from the \emph{natural-language} specification in \Cref{fig:triple}(a) to the \emph{formal} specification in \Cref{fig:triple}(b)?
Simply stating the rules of I/O grammars to an LLM and asking it to produce the grammar directly does not work well.
This is due to the lack of formal specifications to learn from---I/O grammars are a recent invention, and there are no large corpora (yet) of such grammars (or similar protocol specs) available for training.
Consequently, we have to follow a more complex approach, shown in \Cref{fig:pipeline}, making use of a protocol \emph{implementation} to refine the specification:

\begin{enumerate}
\item We use specific tools to extract \emph{protocol elements} from the RFC text, including messages, states, and transitions.
\item We employ an LLM to generate an \emph{initial I/O grammar} based on the extracted protocol elements.
\item We use a \emph{grammar-based fuzzer} to generate tests from the grammar and run them against a \emph{protocol implementation}.
\item We then analyze the \emph{execution traces} of these tests, including successes, failures, and timeouts to identify discrepancies between the grammar and the implementation.
\item These discrepancies are fed back to the LLM, which \emph{repairs} the grammar, preserving traceability to the original RFC.
\item This synthesis-and-repair cycle can be repeated until the grammar is of sufficient quality.
\end{enumerate}

This cycle is effective.
The I/O grammar in \Cref{fig:triple}(b) was automatically produced by our \AUTOSPEC prototype implementation\footnote{\AUTOSPEC = ``\ul{A}uthor \ul{U}tterances \ul{to} \ul{Spec}ifications''} after 3 synthesis-and-repair iterations, and now allows testing real-world POP3 client and server implementations effectively.

We have evaluated \AUTOSPEC on five \emph{internet protocols} (SMTP, POP3, IMAP, FTP, and ManageSieve).
The formal specifications produced by \AUTOSPEC cover average 92.8\% of client and 80.2\% of server message types, and achieves 81.5\% message acceptance across diverse, real-world systems. We see these results as highly promising steps towards full automatic formalization of natural-language specifications for comprehensive test generation.

We expect that once we have obtained sufficiently many mappings between natural-language specifications and formal specifications, we will be able to further train LLMs to formalize specifications automatically even without the need for an implementation in the loop.
This will enable a seamless integration of natural language requirements into the testing process, including evolving and specializing formal specifications towards new natural language requirements.

In summary, our contributions are as follows:

\begin{enumerate}
\item \textbf{An end-to-end system that turns semi-structured natural language specifications into executable formal specifications.}
To the best of our knowledge, ours is the first system to automatically produce formal protocol specifications from natural-language documents, enabling comprehensive automated test generation.
The system checks protocol behavior, not just message formats, and works without protocol-specific tuning.
\item \textbf{An iterative synthesis-and-repair cycle that leverages implementation feedback to refine and repair protocol specifications.}
To the best of our knowledge, ours is the first system to automatically refine and repair protocol specifications based on test feedback from implementations.
\end{enumerate}

In its evaluation on five Internet protocols, our \AUTOSPEC prototype demonstrates both contributions, achieving high coverage of protocol elements and thus effective testing of real-world systems.
Given the abundance of natural-language specifications in practice, this is an important step towards fully automatic formalization of specifications and much more comprehensive test generation.

\section{Background}
\label{sec:background}

\subsection{Testing Protocols}

Generating protocol tests faces two main challenges:
\begin{enumerate*}[label=(\arabic*)]
 \item Generators need to produce inputs that are \emph{valid} and \emph{meaningful} enough to trigger diverse behaviors.
 \item Each test requires an \emph{oracle} that can automatically decide if the system's response is correct~\cite{Barr2015TheOP,grammarwhitefuzz}. This becomes even harder for network protocols.
 \end{enumerate*}

On the wire, a protocol packet is a byte string defined by the standard. Fields may be fixed-width integers, length-prefixed lists, or tagged-union payloads, carrying constraints such as length agreement, checksums, or cross-field equalities, which must hold for the packet to be accepted.

Protocol validity is also temporal and role-dependent. Protocols advance through finite-state machines (FSMs) that define what each peer can send or accept next.

\subsection{Specifying Protocols}

Protocol tests are typically derived \emph{manually} from \emph{natural-language specifications.}
\emph{Request for Comments} documents (RFCs)~\cite{rfc-editor} are the accepted reference for Internet protocols. They combine three elements:  
\begin{enumerate*}[label=(\arabic*)]
  \item normative \emph{keywords} such as \texttt{MUST}, \texttt{SHOULD}, and \texttt{MAY} that define mandatory, recommended, or optional behavior;  
  \item machine-readable \emph{fragments}---grammar fragments, C-style structures, and packet diagrams that define exact encodings, field sizes, and byte order;  
  \item semantic \emph{conditions} such as cross-field invariants and temporal constraints, often described in prose or footnotes~\cite{RFC7322}.
\end{enumerate*}

Together, these elements make RFCs a strong basis for testing.
Any compliant implementation must satisfy all three, which reduces the blind spots of purely differential techniques~\cite{rfcoracle}.

Written in English, RFCs differ in style and vocabulary, rely on implicit domain knowledge, and scatter information across sections, examples, and tables.
Still, they form a \emph{gold standard} for protocol specifications, as they are precisely written publicly vetted.

Note we do not use proprietary Volkswagen specifications in this paper due to confidentiality and system maturity; instead, we bootstrap on public, well-specified protocols, which allows us to measure faithfulness and completeness against known references while developing the methodology for future industrial deployment.

\subsection{Fuzzing Protocols}

In the absence of a formal specification, one can still make use of \emph{implementations} to guide automated test generation.
In the past years, \emph{fuzzing}---generating random inputs to test a system---has become the standard technique for robustness testing.
Coverage-guided mutation fuzzers such as AFL~\cite{aflwhitepaper} and libFuzzer~\cite{Serebryany2016ContinuousFW} work well when random mutations can still produce valid inputs.
However, protocol inputs rarely have this property.
Messages are structured and constraint-heavy, so blind mutation is often rejected early and fails to reach deeper states~\cite{pham2021smart}.
Hence, uniquely relying on coverage of a given implementation places challenges for modern protocol fuzzers:
\begin{enumerate*}
\item heavy dependence on seeds;
\item unknown message structure, which blocks structural mutations; and
\item an unknown state space, which prevents progress to new states~\cite{skyfire,pham2021smart}.
\end{enumerate*}

For our industrial practice, though, the main challenge of using a generic fuzzer is \emph{the lack of an oracle.}
Fuzzers typically assume that bugs are indicated by crashes or hangs.
However, a server may reject a malformed packet without error, or a client may misinterpret a valid but unexpected response.
Testing hardware, such as automotive components, via protocols, requires a precise \emph{oracle} that can check every response for correctness---and in the absence of a formal specification, there is no way to do this fully automatically.

\subsection{Specification-Based Protocol Testing}

To overcome the challenges of coverage-guided fuzzing, one can \emph{formally specify} the protocol syntax, states, transitions, and constraints.
\emph{Grammars,} for instance, are a well-established means to specify the syntax of individual messages: More than 20\% of RFCs include grammars to define message and input formats~\cite{iogrammars}.
Given a comprehensive grammar, a fuzzer can rapidly explore the input space by generating syntactically valid inputs.

Recent works enhance grammars with \emph{constraints}---predicates over grammar elements---to express properties that a context-free grammar cannot capture on own; a constraint such as \texttt{len(<payload>) == uint32(<length\_field>)}, for instance, captures the relationship between two fields in a length-prefixed message~\cite{ZamudioAmaya2025FANDANGO}.

Having an expressive specification for \emph{messages} is a good start, but protocols are more than isolated packets.
The recent concept of \emph{I/O grammars}~\cite{iogrammars} addresses this by having grammars express entire \emph{interactions} between multiple parties.
Here, nonterminals can be prefixed with \emph{party tags} (e.g., \texttt{Client:} or \texttt{Server:}) to indicate who sends what and when.

The POP3 I/O grammar in \Cref{fig:triple}(b), for instance, defines interactions such as \texttt{<Client:USER>}, indicating that the POP3 client issues a \texttt{USER} command, expecting \texttt{Server:user\_response}---a user response from the server.

As it specifies entire interactions, an I/O grammar can be used to generate tests for both clients and servers, checking that each party behaves correctly in every state and transition.
Adding constraints helps to narrow down the interaction space as well as checking that the observed interaction is actually correct.
These properties made us choose I/O grammars as protocol specification formalism over less expressive alternatives~\cite{PeachFuzzer}.

The problem with specification-based test generation is \emph{cost.}
Specifications require human time and expertise to write and maintain.

Under-specification weakens tests, while over-approximation wastes fuzzing budget~\cite{pham2021smart,superion}.
RFCs make this even harder: they are long, heavily cross-referenced, and encode stateful, cross-message constraints.
The one good news in this context is that once a protocol is defined, it tends to be very stable; hence, a formal protocol specification can be used for years to come.

\subsection{Specification Mining with LLMs}

A tempting means to obtain a protocol specification or grammar could be to ``just ask an LLM''.
Indeed, large language models (LLMs) have shown strong results in following instructions and generating text. They can even translate natural-language specifications into executable artifacts with little or no extra training~\cite{llmtext1,llmtext2}.
Protocol standards are published as RFCs, with behavior described in prose and examples.
This makes LLMs well-suited to process RFCs and produce machine-readable representations for testing~\cite{PROSPER}.

However, naïve end-to-end prompting over long documents such as RFCs yields incomplete, misaligned, and non-executable artifacts with missed edge cases and poor traceability~\cite{llmHall}; risks that increase with document length and variability.
Even small hallucinations in constraints can flip verdicts and misdiagnose implementations~\cite{MengMBR24}.

To address these issues, within protocol testing, a growing line of RFC-grounded approaches parses standards into \emph{structured artifacts} that can drive conformance checks. Zheng et al.~\cite{zheng2025validating} segment RFCs and retain paragraph-level anchors to improve traceability and auditable rule reconstruction; PROSPER~\cite{PROSPER} and rfc2fsm~\cite{rfc2fsm} leverage LLMs to lift RFC prose into message syntax, constraints, or state machines.
Complementing these, HDiff~\cite{HDiff} applies NLP to extract ABNF rules from RFCs and build test cases, but---owing to NLP uncertainty---requires manual verification of all ABNF rules before experiments, introducing substantial human effort.
Sage~\cite{rfcsemi} uses classical NLP to surface ambiguities/under-specification and, once clarified by authors, synthesizes interoperable protocol code.

Collectively, these works demonstrate the feasibility of deriving protocol artifacts from standards text, as we do.
However, most target a \emph{single layer} (syntax or FSM) and/or offer limited support for cross-message bindings and temporal guards.
Most importantly, they universally require \emph{manual verification} and lack execution-guided repair toward automatically obtaining a fully executable, comprehensive session-level specification---which is what we aim for, and what we contribute in this paper.

 \begin{figure}[t]
  \centering
  \includegraphics[width=\linewidth]{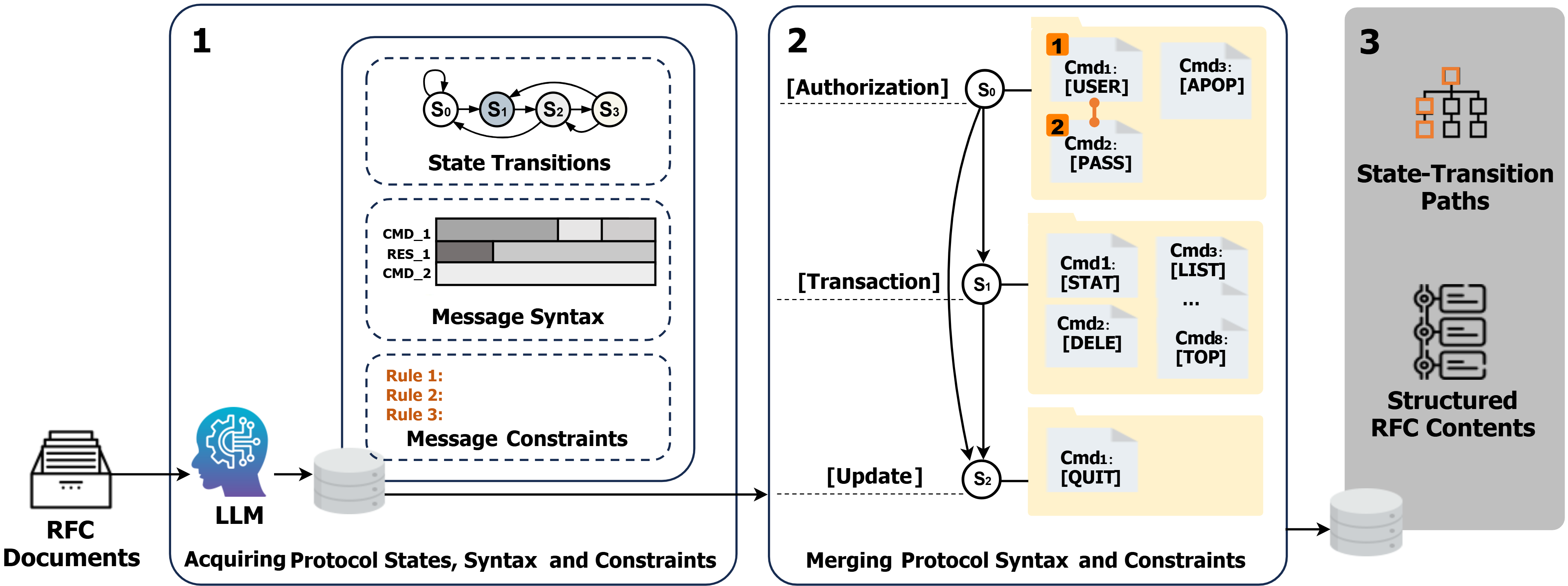}
   \caption[RFC preprocessing pipeline.]{RFC preprocessing pipeline. From the raw RFC text, an LLM (1) extracts three element types: state transitions, message syntax, and message constraints. A merger (2) groups commands by state and links commands by their dependencies. The output (3) comprises state-transition paths and structured RFC contents for I/O grammar synthesis.}
  \label{fig:preprocess}
  
\end{figure}

\section{Preprocessing RFCs}
\label{sec:preprocessing}

Let us now describe our \emph{methodology,} starting with \emph{RFC preprocessing.}
RFCs interleave prose, examples, and formal constraints, so mining I/O grammars or FSMs directly from raw text is brittle. As shown in \Cref{fig:preprocess}, we run one LLM pass per section to extract protocol states, transitions, message syntax, and constraints, then fuse the section outputs into a state-transition \emph{multigraph.} Unlike DocTree-centric approaches~\cite{zheng2025validating}, we 
retain only section IDs and paragraph indices for traceability.

\iflong
\else
At this point, preprocessing is required to extract relevant information from RFCs. Eventually, a sufficiently large \emph{corpus} of mappings between natural-language specifications and formal specifications may enable end-to-end translation without preprocessing.
\fi

\subsection{Extracting States and Transitions}

We use a three-step, LLM-orchestrated pipeline to extract state and transition knowledge from RFCs, decoupling classification, extraction, and synthesis.
This improves robustness: the classifier handles categorization, the extractor handles fine-grained lifting, and the synthesizer performs graph-level reasoning.

\subsubsection{Section Filtering.} The model inspects each section, filters out irrelevant text, and assigns a semantic label (state machine, overview, example, other) plus an action (extract, copy, summarize), so only rule-bearing sections proceed.

We use a structured prompt with four blocks: Objective, Requirements, and explicit Input/Output formats (\Cref{fig:prompt2}).
The model is instructed to act as a domain-specific RFC expert. It classifies a section and returns JSON that follows a four-way schema. We impose constraints ($\leq$30-token summary, no extra keys) to make outputs deterministic. For traceability, the classifier echoes the section ID and the paragraph indices that triggered each decision. The output is a list of \{\texttt{section\_id}, \texttt{label}, \texttt{action}, \texttt{summary}\} records. Sections tagged \texttt{extract} move to the next stage; \texttt{copy} and \texttt{summarize} are kept as context without incurring extraction cost.

\subsubsection{Structured Extraction.} For sections marked \texttt{extract}, the model lifts states, commands, transitions, and syntax/constraint rules into normalized JSON for automated downstream use.

\begin{figure}[t] 
  \centering
  \includegraphics[width=1\linewidth]{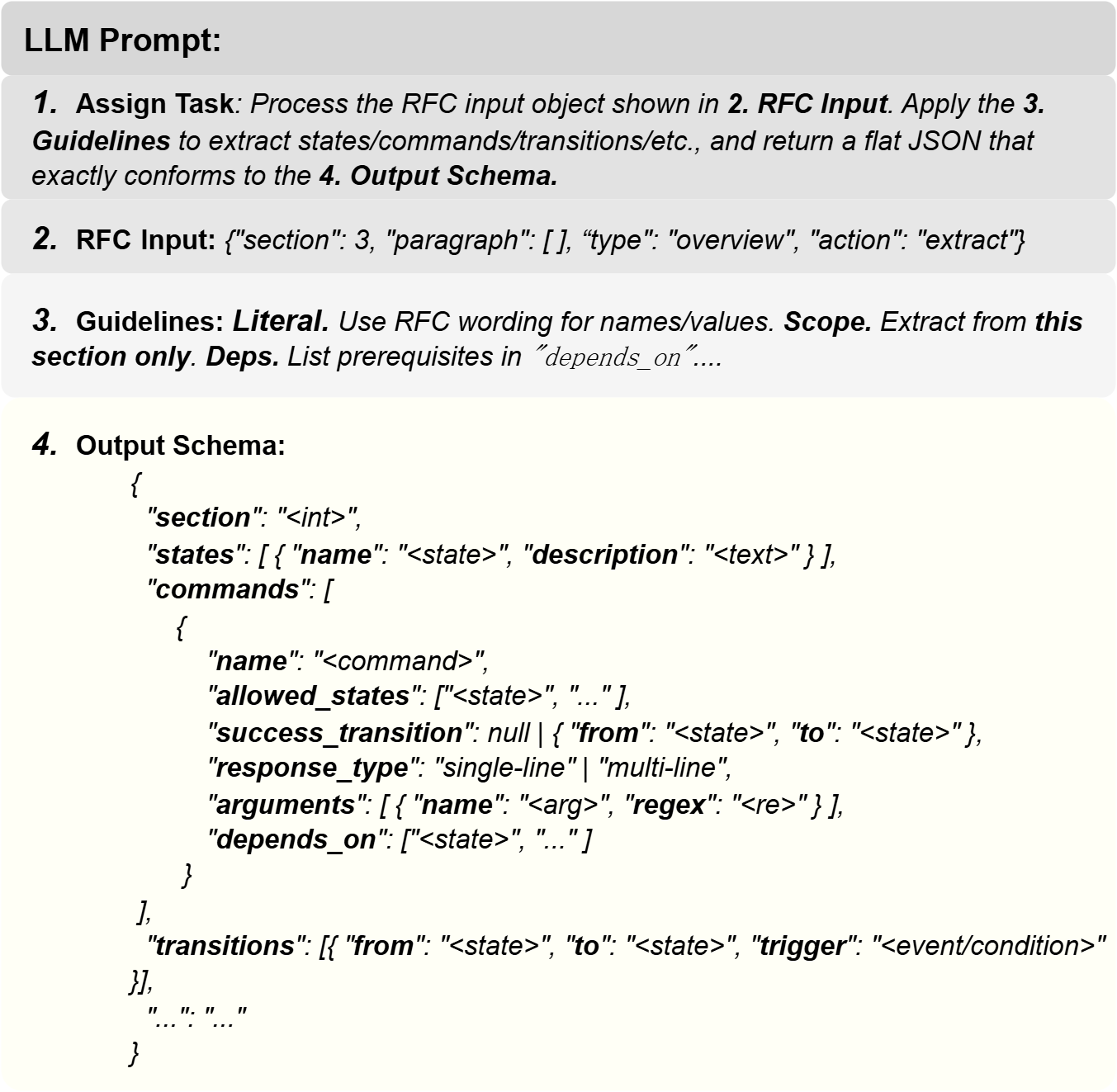}
  \caption{Prompt for states/commands/transitions extraction. 
}

  \label{fig:prompt2}

\end{figure}

\begin{figure}[h] 
  \centering
  \includegraphics[width=1\linewidth]{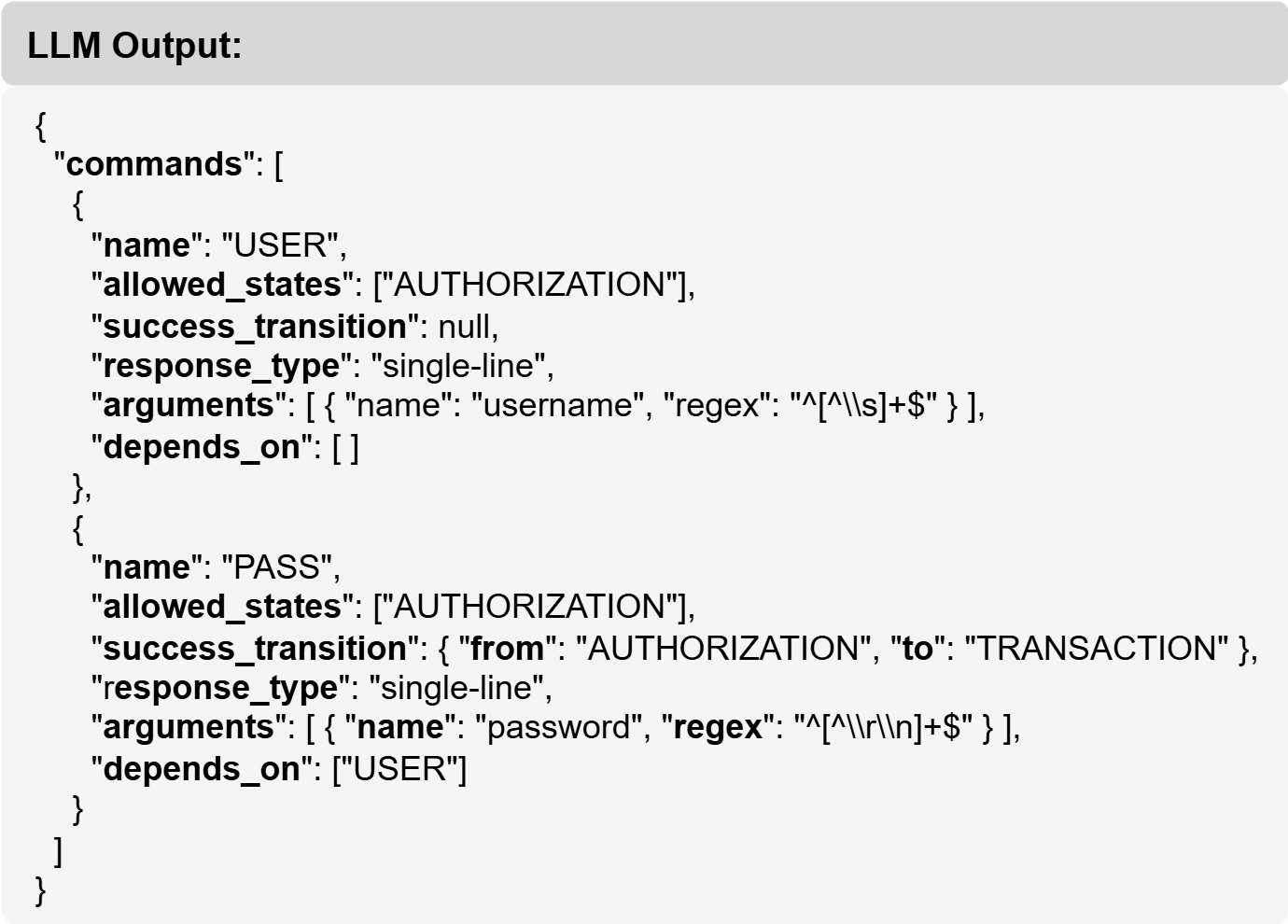}
  
  \caption{LLM output for POP3 \texttt{USER}/\texttt{PASS} (\Cref{fig:prompt2}).}

  \label{fig:llmexampl}
\end{figure}

Sections with \texttt{action}=\texttt{extract} trigger a second prompt that projects natural-language constraints into a JSON fragment; see \Cref{fig:prompt2} for the schema and \Cref{fig:llmexampl} for a POP3 extraction example.
Missing pieces are filled as \texttt{null} to keep the output total and machine-checkable. Each fragment is validated at run time by a lightweight \textit{Pydantic}~\cite{pydantic} model, which rejects missing fields, type mismatches, and out-of-range values. For example, Section~5 of RFC~1939 yields a fragment that describes the \textsc{TRANSACTION} state, six POP3 commands, and a single \textsc{QUIT}$\rightarrow$\textsc{UPDATE} transition. The fragment passes validation without manual edits and can be merged as is. If validation fails, we discard the instance and re-issue the prompt once; this resolves most cases.

\subsubsection{Multigraph Synthesis.}
Each JSON fragment becomes a section-local graph and is then merged into a unified state-transition multigraph. Nodes denote states, commands, and responses; edges encode transitions and command dependencies.

The Multigraph Synthesis module proposes a multigraph from each validated fragment produced in Structured Extraction. This stage continues to use the LLM, now to consolidate each section’s validated JSON fragment into a typed, section-local \emph{micrograph}. The LLM assists with (i) \textit{name harmonization} (e.g., resolving aliases and capitalization to canonical forms), (ii) \textit{intra-section consolidation} (deduplicating near-duplicates and merging synonymous rules), and (iii) \textit{edge completion} (recovering implied transitions or dependencies mentioned across sentences). These steps reduce conflicts in the subsequent global merge.

We type nodes as \emph{states}, \emph{commands}, and (optionally) \emph{responses}. We type edges as:
\emph{invokes} (state$\rightarrow$command), 
\emph{yields} (command$\rightarrow$state/\allowbreak response), and 
\emph{dependencies} between commands (later normalized into \emph{requires}/\emph{enables}).
For reproducibility, we do not accept the LLM output verbatim. We apply a deterministic \emph{acceptance filter}
that (i) enforces node/edge typing, (ii) requires provenance anchors (section ID and paragraph indices) for \emph{both} nodes and edges, and (iii) resolves duplicates by precedence (normative text $>$ examples/overviews). Only accepted section-local micrographs are forwarded to the global merger.

\subsection{Merging States and Transitions}

After extracting states and transitions, we consolidate (i) the per-section JSON fragments and (ii) the accepted section-local micrographs into a protocol-level, state-transition multigraph. Unlike Multigraph Synthesis, this merger is fully deterministic and rule-based; it preserves section provenance and ensures reproducibility.

\subsubsection{Deterministic merger}
We first \emph{normalize labels} (for instance, \texttt{USER}/\texttt{User}/\texttt{user} $\rightarrow$\texttt{USER}; \texttt{Auth}$\rightarrow$\texttt{AUTHORIZATION}) and \emph{canonicalize edge types} by mapping \textit{dependencies} into \texttt{requires}/\texttt{enables}. We then take the union of nodes and edges while preserving provenance. Next, we propagate constraints: section-level pre/postconditions are lifted onto the relevant states and edges, combined by intersection when compatible, and flagged when \iflong
\else
contradictory.
\fi

The result is an FSM-like graph
$G=\langle S, C, E, \Phi\rangle$
with states $S$, commands $C$, typed edges $E$ (invokes, yields, requires/enables), and a global constraint store $\Phi$. This rule-based consolidation avoids hallucinations and provides a reproducible baseline for testing.

\subsubsection{Minimal transition paths}
\label{sec:mtps}
From the consolidated protocol multigraph, we compute a set of \emph{minimal transition paths (MTPs)}. 
Each MTP represents the shortest sequence of states and commands required to achieve a target behavior, respecting all dependencies and preconditions. 
These paths serve as compact, context-rich cues for the Grammar Generation Agent, guiding the synthesis process towards specific protocol interactions.
In details, from the initial state(s) $S_0$, we compute, for each target command or transition, a \emph{minimal state-transition path}: the shortest edge sequence that (i) respects all dependencies, (ii) satisfies accumulated preconditions, and (iii) establishes the required postcondition in the destination state. 
For POP3, this yields paths such as
\[
S_0 \xrightarrow{\texttt{USER}} S_0
\xrightarrow{\texttt{PASS}} S_1
\xrightarrow{\texttt{STAT/LIST/DELE}} S_1
\xrightarrow{\texttt{QUIT}} S_2,
\]

These paths act as \emph{retrieval keys}: each path pinpoints the RFC clauses (and their constraints/grammars) needed to exercise a target behavior, focusing the agent loop on the smallest valid context.

\subsubsection{Outputs}
The merger generates two artifacts (rightmost panel of \Cref{fig:preprocess}):
\begin{enumerate*}[label=(\arabic*)]
  \item \emph{Structured RFC Contents}: a normalized index of states, commands, constraints, and their section provenance;
  \item \emph{State-Transition Paths}: a compact set of paths that capture valid protocol progressions with prerequisites and postconditions resolved.
\end{enumerate*}
These artifacts drive the next stage: test grammars and messages are synthesized conditioned on minimal paths; inconsistencies are detected; and the consolidated model is iteratively repaired.

\section{Synthesizing I/O Grammars}
\label{sec:synthesizing}

The \emph{Generator Agent} turns the preprocessed RFC specifications from \Cref{sec:preprocessing} into an executable I/O grammar.

\subsection{Synthesis}

We follow a five-step pipeline that progressively moves from text to a normalized, engine-ready grammar:

\subsubsection{Path$\rightarrow$FSM skeleton}
Minimal transition paths (MTPs; \Cref{sec:mtps}) serve as the top-level skeleton. The agent converts each path into \emph{state-command-transition} triples and builds a minimal FSM outline, determining the nonterminals and the order of exchanges.

\subsubsection{Evidence alignment from RFC excerpts}
We align the retrieved RFC text (normative clauses and any sectioned code/message listings) to the FSM outline. The agent fills the skeleton by emitting draft productions and mapping concrete request/response lines to terminals and role aliases (for instance, \texttt{<Client\:command>}, \texttt{<Server\:response>}). 

\subsubsection{Constraint enrichment}
The agent attaches \emph{constraints} to the relevant nonterminals and fields as \texttt{where} clauses. Supported forms include value ranges, length agreements, cross-field equality, and functional/derivation constraints.

\subsubsection{Normalization and output}
The agent produces the final I/O grammar that (i) instantiates <start> with the MTP-ordered exchanges, (ii) groups productions under their nonterminals, (iii) deduplicates repeated command forms, and (iv) includes a <terminals> block for literal/lexeme definitions.

\subsubsection{Patch integration from fixer outputs}
When the repair stage produces patches, the generator applies them to the current grammar: (i) update the targeted productions and associated \texttt{where} constraints using the provided locations; (ii) maintain a minimal diff while preserving provenance (RFC\_ID/section/paragraph); (iii) re-run deduplication and normalization (grouping, alias consistency, and \texttt{<terminals>} completeness); and (iv) enforce schema/model checks (Pydantic schema) and MTP generatability checks. The revised grammar is then re-executed against the SUT, and the results are fed back to the Repair Agent.

\subsection{Prompt Design and Guardrails}

\begin{figure}[t]
  \centering
  \includegraphics[width=1\linewidth]{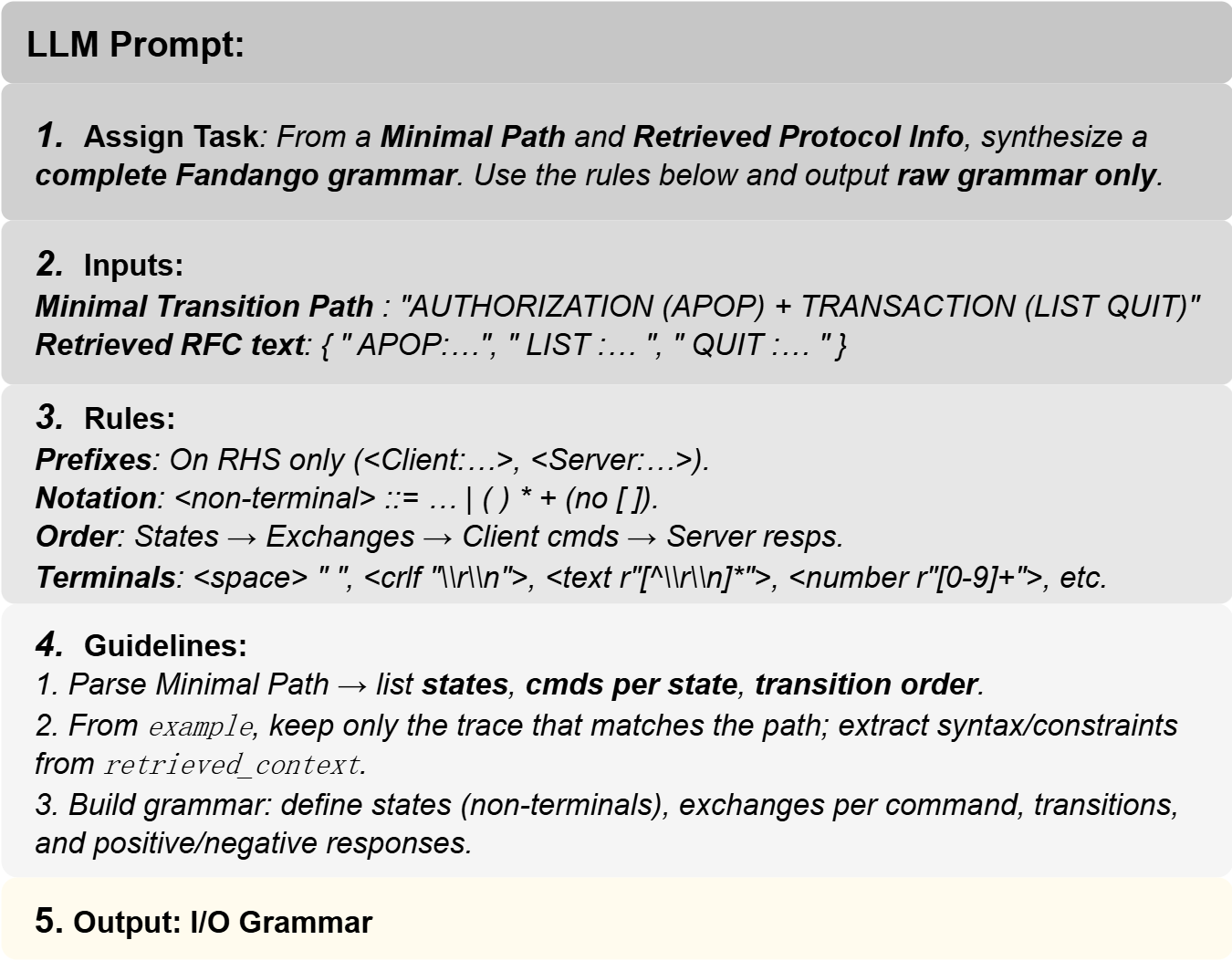}
  \caption{Prompt for I/O Grammar Generator Agent.}
  
  \label{fig:prompts21}
\end{figure}

The generator prompt (\Cref{fig:prompts21}) specifies: (i) the \emph{Task} (``Generate I/O Grammar from MTPs and retrieved \{\texttt{RFC\_ID}\} text''), (ii) a target \emph{Grammar scaffold} (required nonterminals/blocks), (iii) a concise \emph{Workflow} (MTP$\rightarrow$FSM, draft from retrieved text, emit constraints), and (iv) strict \emph{output rules} (``grammar only; no commentary''). Guardrails enforce literal symbol names, section-local scope, normative-over-example precedence, provenance quoting, and schema checks. Each output is validated by a Pydantic schema; invalid outputs are rejected and the prompt is re-issued once.

\begin{figure}[t]
  \centering
  \includegraphics[width=1\linewidth]{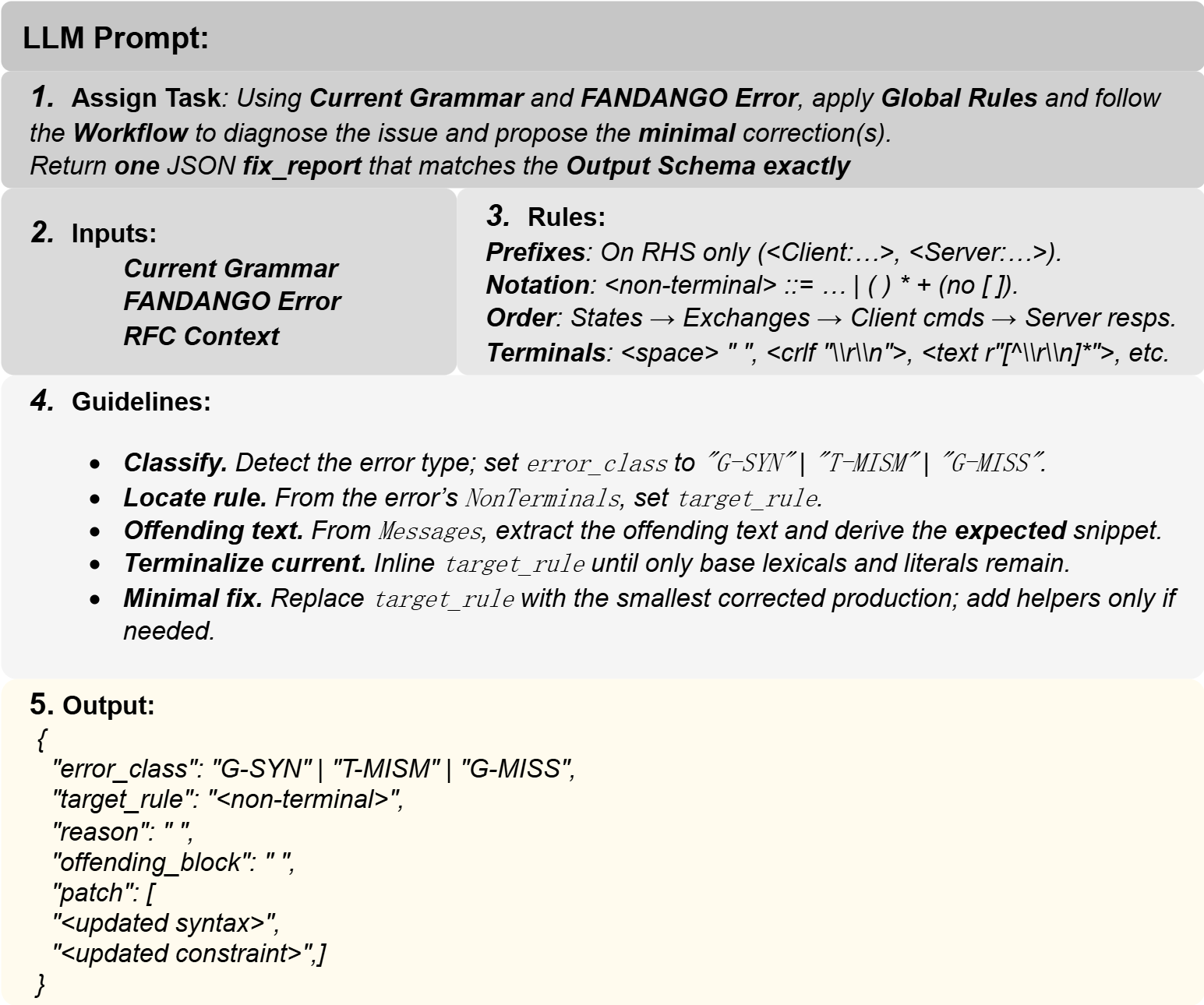}
  \caption{Prompt for I/O Grammar Fixer Agent.}
  \label{fig:fix_prompt}
\end{figure}

\section{Repairing and Refining I/O Grammars}
\label{sec:repairing}

The \emph{Grammar Repair Agent} receives the I/O grammar from \Cref{sec:synthesizing} together with execution outputs from the grammar-based fuzzer and SUT (parser errors and request-response traces in \Cref{fig:pipeline}). It turns runtime evidence into the smallest patch that the generator applies to synthesize a revised I/O grammar (see \Cref{fig:fix_prompt}).

\subsection{Workflow}

The agent follows a five-step workflow:

\subsubsection{Classification}
The agent inspects the error line and trace, and maps it to an error class (or None). It records the \texttt{error\_class} and a short \texttt{reason}.

\subsubsection{Target localization and evidence capture}
Using the parser stack and execution trace, the agent identifies the nonterminal (\texttt{target\_rule}) to edit and the concrete input fragment. 

\subsubsection{Diff reduction}
The agent inlines the \texttt{target\_rule} at the failure site to obtain a \texttt{current\_snippet}. Abstract the triggering request/response into an \texttt{expected\_snippet}. Reduce both to terminal-level sequences and compute their minimal edit script; this yields a precise delta for patch generation.

\subsubsection{Patch generation}
\iflong
\else
The agent emits a Fandango-compatible patch, rewriting the production, updating constraints, or adding missing alternatives, depending on the error class.
\fi
  Patches follow a \emph{minimal change} policy and preserve canonical symbol names.
\subsubsection{Structured output} The agent returns a JSON fix report (\Cref{fig:pop3-top-syntax}) with error class, location, rationale, and patch entries. The report goes into the \emph{I/O Grammar generator} as guidance for regeneration, and the MTPs that exercise the patched rules are re-tested.

\subsection{Result: A Refined I/O Grammar}

The loop stops when schema checks pass and all MTPs can be generated, or when a repair budget is exhausted.
At this point, we have a refined I/O grammar that is (i) schema-valid, (ii) MTP-generatable, and (iii) tested against real implementations. This grammar can then drive grammar-based fuzzing for conformance testing.

\section{Implementation}
\label{sec:implementation}

\AUTOSPEC is a fully automated pipeline that ingests RFC HTML, normalizes each section into JSON, and drives a two-stage LLM workflow to synthesize and repair executable I/O grammars, which are then fuzzed against real servers with the grammar-based fuzzer Fandango~\cite{ZamudioAmaya2025FANDANGO}. 
A crawler cleans RFC pages---removing boilerplate, tables of contents, and footers while preserving section IDs and paragraph indices---and emits one JSON file per section. 
A section-wise LLM

classifies sections and extracts states, commands, and constraints to build a state-transition multigraph. 
Given this multigraph and the RFC text, Stage~2 generates an I/O grammar and iteratively patches it using execution feedback.

To avoid cross-talk between roles, each agent runs in its own sandbox.

All model outputs are schema-constrained JSON and are validated locally (e.g., via Pydantic); invalid outputs trigger bounded retries with minimal edits. 
We cache retrieval results and model responses and apply client-side rate limiting to respect API quotas. 
The orchestrator is CPU-only and communicates with cloud LLM backends; systems under test (SUTs) run in containers or on sockets.
Unless noted otherwise, we use a GPT-4-class model~\cite{gpt4} with fixed decoding settings (\texttt{temperature}=0, \texttt{top\_p}=1.0, and no \texttt{stop} sequences) for reproducibility. We use a low temperature for determinism—\texttt{temperature}=0.1 for generation and \texttt{temperature}=0.0 for lightweight classification—and allow up to \texttt{max\_tokens}=4000 for non-streaming long generations (grammar synthesis); all other calls use the library default cap.
We keep structured logs (requests, responses, traces, diffs, patches) and store artifacts under a stable layout (\texttt{data/RFC/}, \texttt{output/flows/}, \texttt{output/grammars/}, \texttt{output/runs/}). 
Models (GPT-4-class) and protocol targets (POP3, IMAP, ManageSieve, SMTP, FTP) are pluggable via configuration settings.

\begin{table*}[t]
  \footnotesize
  \renewcommand{\arraystretch}{1.1} 
  \setlength{\tabcolsep}{4pt}
  \caption{Protocol Dataset and Ground-Truth I/O Grammar}
  \label{tab:protocol-data}
  \rowcolors{2}{gray!20}{white}
  \begin{tabularx}{\linewidth}{@{}
    l l r | l l l | r r r r r r r r @{}}

    \rowcolor{Goldenrod} 
    \multicolumn{3}{@{}c|}{\textbf{Protocol}} &
    \multicolumn{3}{c|}{\textbf{Implementation}} &
    \multicolumn{8}{c@{}}{\textbf{Ground-Truth I/O Grammar}}\\

    \rowcolor{Goldenrod!50}
    & & & & & \textbf{Tested} & \textbf{Msg Types} & \textbf{Nonterminal} & \textbf{Indep.} & \textbf{Dep.} & \textbf{Msg} & \textbf{Trace} & & \\
    \rowcolor{Goldenrod!50}
    \textbf{Name} & \textbf{RFC} & \textbf{Pages} & \textbf{Name} & \textbf{Port} & \textbf{Party} &
    \textbf{(Total; C/S)} & \textbf{Name/Type} & \textbf{Constr.} & \textbf{Constr.} &
    \textbf{Acceptance} & \textbf{Acceptance} & \textbf{LoC} & \textbf{Time} \\

    POP3        & 1939 & 23  & Dovecot 2.3.19.1  & 110  & server & 24 (14/9) & 14/5 & 12 & 1 & 13/14(92.9\%) & 9/10(90.0\%) & 295 & 3.5h \\
    SMTP        & 5321 & 95  & Postfix 3.7.11    & 25   & server & 23 (11/11) & 18/6 & 8  & 0 & 8/11(72.7\%)  & 4/6(66.7\%)  & 373 & 9.5h \\
    IMAP        & 3501 & 108 & Dovecot 2.3.19.1  & 143  & server & 48 (30/17) & 26/5 & 14 & 3 & 29/30(96.7\%) & 15/16(93.8\%) & 676 & 14h \\
    FTP         & 959  & 69  & vsFTPd 1.3.7a     & 21   & server & 43 (31/15) & 12/5 & 5  & 3 & 28/31(90.3\%) & 8/11(72.7\%)  & 576 & 5.5h \\
    ManageSieve & 5804 & 49  & Pigeonhole 0.5.19 & 4190 & server & 21 (14/6)  & 9/5  & 12 & 1 & 12/14(85.7\%) & 9/11(81.8\%)  & 449 & 5h \\
  \end{tabularx}
  \vspace{0.3em}
  \captionsetup{font=footnotesize}
  \caption*{\footnotesize\textit{Notes.} Services run in containers:
  POP3/IMAP/SMTP/ManageSieve via \texttt{docker-mailserver:latest} (Debian 12, OpenSSL 3.0.15); FTP results cover the control-\\channel only; data-channel semantics and payloads are out of scope; Msg types are reported as  \emph{Total; C=client requests, S=server replies}; Indep./Dep. Constr. \\ denote independent/dependent constraints.}
\end{table*}

\begingroup
\colorlet{Goldenrod}{RoyalBlue!20!white}
\begin{table}[t]
  \footnotesize
  \renewcommand{\arraystretch}{1.1}
  \setlength{\tabcolsep}{3.5pt}
  \caption{RQ1: I/O Grammar Extraction: Recall.}
  \label{tab:rq1recall}

  \rowcolors{2}{gray!20}{white}
  \begin{tabularx}{\linewidth}{@{}
      l
      >{\centering\arraybackslash}X
      >{\centering\arraybackslash}X
      >{\centering\arraybackslash}X
      >{\centering\arraybackslash}X
    @{}}

    \rowcolor{Goldenrod}
    \multicolumn{1}{@{}c}{\textbf{Protocol}} &
    \textbf{Client Msg Type} &
    \textbf{Server Msg Type} &
    \textbf{Indep.\ Constr.} &
    \textbf{Dep.\ Constr.} \\

    POP3        & 14/14(100.0\%) & 9/9(100.0\%)   & 10/12(83.3\%) & 0/1(0.0\%) \\
    SMTP        & 11/11(100.0\%) & 7/11(63.6\%)   & 0/8(0)        & 0/0(-)     \\
    IMAP        & 25/30(83.3\%)  & 12/17(70.6\%)  & 0/14(0)       & 0/3(0\%)     \\
    FTP         & 25/31(80.6\%)             & 10/15(66.7\%)              & 0/5(0\%)          & 0/3(0\%)      \\
    ManageSieve & 14/14(100.0\%)             & 6/6(100.0\%)             & 0/2(0\%)          & 0/1(0\%)      \\
    \textbf{Average} & 92.8\% & 80.2\% & 16.7\% & 0.0\% \\
  \end{tabularx}

  \vspace{0.25em}
  \captionsetup{font=footnotesize}
  
\end{table}

\colorlet{Goldenrod}{RoyalBlue!20!white}
\begin{table}[t]
  \footnotesize
  \renewcommand{\arraystretch}{1.1}
  \setlength{\tabcolsep}{3.5pt}
  \caption{RQ1: I/O Grammar Extraction: Precision.}
  \label{tab:rq1precision}

  \rowcolors{2}{gray!20}{white}
  \begin{tabularx}{\linewidth}{@{} 
      l
      >{\centering\arraybackslash}X
      >{\centering\arraybackslash}X
      >{\centering\arraybackslash}X
      >{\centering\arraybackslash}X
    @{}}

    \rowcolor{Goldenrod}
    \multicolumn{1}{@{}c}{\textbf{Protocol}} &
    \textbf{Client Msg Type} &
    \textbf{Server Msg Type} &
    \textbf{Indep.\ Constr.} &
    \textbf{Dep.\ Constr.} \\

    POP3        & 14/14(100.0\%) & 9/9(100.0\%)   & 11/13(84.6\%) & 0/0(-) \\
    SMTP        & 11/11(100.0\%) & 7/7(100.0\%)   & 0/0(-)        & 0/0(-)    \\
    IMAP        & 25/25(100.0\%) & 12/12(100.0\%) & 0/0(-)        & 0/0(-) \\
    FTP         & 25/26(96.2\%)             & 10/14(71.4\%)             & 0/0(-)          & 0/0(-)  \\
    ManageSieve & 14/14(100.0\%)             & 6/6(100.0\%)              & 0/2(0\%)          & 0/1(0\%)  \\
    \textbf{Average} & 99.2\% & 94.3\% & 42.3\% & 0.0\% \\
  \end{tabularx}

  \vspace{0.25em}
  \captionsetup{font=footnotesize}
  
\end{table}

\section{Evaluation}
\label{sec:evaluation}

We evaluate \AUTOSPEC along two axes: (i) diversity w.r.t.\ expert-authored \emph{golden I/O grammars and (ii) precision on \emph{real systems}.}

\begin{description}
  \item[RQ1: Diversity.] \emph{How much of the protocol do the generated \emph{I/O grammars} cover relative to a golden specification?}
  We compare coverage of message types, fields, and constraints against expert-authored golden I/O grammars.
  \item[RQ2: Precision.] \emph{What precision do the generated I/O grammars achieve on real systems?}
  We measure the precision of I/O grammars by deriving test cases on real systems and measure the percentage of accepted messages and traces.
  \item[RQ3: Repair impact.] \emph{What is the impact of the execution-guided repair loop?} We perform an ablation by removing the 
  loop.
  \item[RQ4: Naïve prompting.] \emph{Can an LLM, without RFC input, produce useful I/O grammars?}
  We compare against a naïve LLM baseline elicited without RFC context.
\end{description}

\subsection{Measures and Ground Truth}
\label{sec:dataset-groundtruth}

We evaluate \AUTOSPEC on five Internet protocols (\emph{POP3}, \emph{IMAP}, \emph{ManageSieve}, \emph{FTP}, and \emph{SMTP}) (\Cref{tab:protocol-data})
against ``golden grammars'' written by one of us as protocol experts.
We execute each golden grammar once
\iflong
\else
for sanity checks.
\fi
The following measures, listed in \Cref{tab:protocol-data}, characterize the golden grammars:
\begin{description}
  \item[Message Types.]
  The distinct on-wire formats a protocol uses for different kinds of messages. We count both \emph{requests} (client commands)
  and \emph{responses} (server reply-line formats), collapsing purely lexical aliases.

  \item[Nonterminal Names.]
  The names of individual data fields within a given message type (e.g., \texttt{number}, \texttt{text}, \texttt{email\_address}).
  We exclude formatting tokens (e.g., \texttt{<space>}, \texttt{<crlf>}) as well as fixed string literals (e.g., \texttt{"EHLO"}).

  \item[Nonterminal Types.]
  The data-type categories and size disciplines of fields (e.g., \emph{byte}, \emph{bit}, \emph{uintN}, \emph{string/atom}, \emph{array/list},
  \emph{struct/sequence}), distinguishing fixed-size from variable-size forms (including those whose length depends on another field).
  A field contributes one \emph{name} and one \emph{type} to the counts.

  \item[Independent Constraints.]
  Restrictions on \emph{single} fields (numeric ranges, enumerations, regex membership, fixed values) that do not reference any
  other field.

  \item[Dependent Constraints.]
  Constraints that relate \emph{multiple} fields or temporal elements (equality, length agreement, checksum relations, 
  ``must follow'' conditions).

  \item[Message Acceptance.]
  When executing a grammar against a real implementation, the fraction of issued \emph{commands} that receive a positive, protocol-conformant response (e.g., POP3 \texttt{+OK}, SMTP 2xx).

  \item[Trace Acceptance.]
    A trace is \emph{accepted} if all steps are accepted and the session reaches the expected terminal state (e.g., POP3: \texttt{USER}\,$\rightarrow$\,\texttt{PASS}\, $\rightarrow$\,\allowbreak \texttt{RETR}\,$\rightarrow$\,\texttt{QUIT}). Rejections, premature disconnect, or response format violations mark the trace as not accepted.

  \item[Lines of Code (LoC).]
  Non-blank, non-comment lines of the golden grammars, as a proxy for specification size.

  \item[Time.]
  The expert’s wall-clock time to author the golden grammar.
\end{description}
In RQ2, we also test the resulting I/O grammars against real implementations of each protocol, listed in \Cref{tab:protocol-data}.

\subsection{RQ1: Diversity}
\label{sec:rq1-diversity}

\emph{How much of the protocol do the grammars cover relative
to a golden specification?}
We evaluate specification coverage against the golden I/O grammars of the five RFC protocols in the full pipeline setting. 

All runs use the same retrieval configuration and fixed LLM decoding hyperparameters; execution is performed against the same SUTs across all conditions.
For each synthesized grammar, we generate tests with at most 7 iterations per MTP per protocol.
We compare the  \emph{synthesized I/O grammars} against the ground truth and report
\emph{precision} and \emph{recall} for each element type (\textsc{Client Msg Type}, \textsc{Server Msg Type}, \textsc{IndepC}, and \textsc{DepC}), using the metrics and abbreviations defined in \Cref{tab:protocol-data}.
Formally, for any element set $E$:
\[
\text{Precision}(E)=\frac{|E_{\text{ours}}\cap E_{\text{gt}}|}{|E_{\text{ours}}|},\qquad
\text{Recall}(E)=\frac{|E_{\text{ours}}\cap E_{\text{gt}}|}{|E_{\text{gt}}|}.
\]
We report per-protocol results in Table~\ref{tab:rq1recall} and ~\ref{tab:rq1precision}. For POP3, the 9 traces cover 14 commands, all 14 were recovered. Per-protocol results appear in Tables~\ref{tab:rq1recall} and \ref{tab:rq1precision}. Under our generate–execute–repair loop, we observe a consistent pattern: \emph{client-side message types} are recovered almost completely across protocols.

\conclusion{\AUTOSPEC-extracted protocol specifications achieve  \textbf{high message type recall} compared to ``golden'' specifications.}

We note, though, that \emph{server-side reply types} lag (notably SMTP/IMAP/FTP). The former are typically enumerated explicitly in RFC ABNF (verbs and arguments); the latter are often described schematically, so the agent captures the reply schema but not every concrete reply type enumerated by the golden. Constraint extraction is weaker: \textsc{IndepC} and especially \textsc{DepC} are predominantly expressed as dispersed normative prose (ranges, conditioned enums, cross-field/temporal relations), leading the agent to mirror surface syntax while under-extracting such semantics.
Consequently, protocols intertwining base syntax with stateful behavior and optional extensions (IMAP tagged/untagged updates; FTP multi-line replies) show lower recall, while protocols with compact exchanges and explicit reply sets (POP3, ManageSieve) approach completeness.

\conclusion{In \AUTOSPEC-extracted protocol specifications, recall of \textbf{server-side replies} is lower, but still high on compact exchanges.}

Let us now turn to \emph{precision,} that is, the fraction of protocol elements in the synthesized grammar that are also in the golden grammar.
We observe \emph{near-perfect precision} on client message types across all \iflong
\else
protocols.
\fi

\conclusion{\AUTOSPEC-extracted protocol specifications achieve  \textbf{high precision} compared to ``golden'' specifications.}

\begingroup
\colorlet{Goldenrod}{RoyalBlue!20!white}
\begin{table*}[t]
  \footnotesize
  \renewcommand{\arraystretch}{1.1}
  \setlength{\tabcolsep}{4pt}
  \caption{RQ2: Execution Precision on Real Implementations}
  \label{tab:rq1-precision}
  \rowcolors{2}{gray!20}{white}
  \begin{tabularx}{\linewidth}{@{} l | r r r | r r r | X @{}}

    \rowcolor{Goldenrod}
    \multicolumn{1}{@{}c|}{\textbf{Protocol}} &
    \multicolumn{3}{c|}{\textbf{Message Acceptance (\%)}} &
    \multicolumn{3}{c|}{\textbf{Trace Acceptance (\%)}} &
    \multicolumn{1}{c@{}}{\textbf{Failure Causes}}\\

    \rowcolor{Goldenrod!50}
    & \textbf{\#Msgs} & \textbf{MA} & \textbf{RtCMA}
    & \textbf{\#Traces} & \textbf{TA} & \textbf{RtCTA}
    & \textbf{Needs\textendash TLS / Impl\textendash Missing / Data\textendash State / Grammar\textendash Bug} \\

    POP3        & 14 & 13/14(92.9\%)   & 92.9/92.9(100.0\%) & 10  & 9/10(90.0\%)   & 90.0/90.0(100.0\%) & Impl\textendash Missing 7.1\% \\
    SMTP        & 11 & 8/11(72.7\%)    & 72.7/72.7(100.0\%) & 14 & 8/14(57.1\%)    & 57.1/66.7(85.6\%) & Impl\textendash Missing 27.3\% \\
    IMAP        & 30 & 27/30(90.0\%)   & 90.0/96.7(93.1\%)  & 24 & 21/24(87.5\%)    & 87.5/93.8(93.3\%)  & Grammar\textendash Bug 6.6\%, Needs\textendash TLS 3.3\% \\
    FTP         & 31 & 25/31(80.6\%)    & 80.6/90.3(89.3\%)  & 22 &16/22(72.7\%)     & 72.7/72.7(100.0\%) & Grammar\textendash Bug 9.7\%, Impl\textendash Missing 9.7\% \\
    ManageSieve & 14 & 10/14(71.4\%)    & 71.4/85.7(83.3\%)  & 13 & 9/13(69.2\%)  & 69.2/81.8(84.6\%) & Grammar\textendash Bug 14.3\%, Data\textendash State 7.1\%, Needs\textendash TLS 7.1\% \\
  \end{tabularx}

  \vspace{0.25em}
  \captionsetup{font=footnotesize}
  \caption*{\footnotesize

    \textit{Notes.} MA/TA are acceptance rates against the golden grammar (GT): per GT command type and per GT canonical trace. 
RtCMA/RtCTA are the corresponding rates over the GT’s \emph{routes-to-canonical} subset (if applicable). 
Failure causes are labeled per non-acceptance event.
  }
\end{table*}

\subsection{RQ2: Precision}
\label{sec:rq2-precision}

\emph{What precision do the generated grammars achieve on real systems?}
Setup, SUTs, and the 7-iteration cap follow RQ1; execution details are provided in \S\ref{sec:implementation}.
\Cref{tab:rq1-precision} reports precision at two granularities, \emph{message acceptance (MA)} and \emph{trace acceptance (TA)}, as introduced in \Cref{sec:dataset-groundtruth}.
Here, \textbf{Msgs} denotes the total number of client commands actually executed (summed over all traces and SUTs) and forms the denominator for MA; \textbf{Traces} denotes the total number of end-to-end sessions executed (across SUTs) and forms the denominator for TA. For POP3, \textbf{13/14} messages are accepted (\textbf{92.9\% MA}) and \textbf{9/10} traces complete (\textbf{90.0\% TA}).

\conclusion{\AUTOSPEC-extracted protocol specifications achieve~high~precision on~real~systems.}
\textbf{RtCMA} (Route-to-Canonical Message Acceptance) and \textbf{RtCTA} (Route-to-Canonical Trace Acceptance) restrict MA/TA to the subset of a golden specification actions that lie on canonical routes; we report them as \emph{overall / canonical} to contrast overall acceptance with acceptance along the GT’s canonical routes.

For SMTP, \emph{RtCTA} exceeds \emph{TA} (66.7\% vs.\ 57.1\%), indicating that most end‑to‑end failures arise off the canonical route, whereas \emph{RtCMA} equals \emph{MA} (both 72.7\%), showing message‑level rejections are not concentrated on canonical commands. IMAP and ManageSieve exhibit the same trend—\emph{RtCTA} improves over \emph{TA} (93.8\% vs.\ 87.5\% and 81.8\% vs.\ 69.2\%)—suggesting the grammars are sound for core flows; remaining drops align with SUT limitations or preconditions (\textsc{Impl–Missing}, \textsc{Needs–TLS}, \textsc{Data–State}) rather than systematic grammar errors.

\conclusion{RtCMA/RtCTA remain high—rejections mainly reflect unmet prerequisites, not synthesis mistakes.}

\subsection{RQ3: Repair Impact}
\label{sec:rq3-repair-impact}

\emph{What is the impact of the execution-guided repair loop?}
We ablate around the Repair Agent to quantify its marginal contribution and cost. We compare two primary configurations under identical RFC set, retriever and SUTs, \textbf{T0 (Text-only)}---generator but no execution feedback; and \textbf{T1 (Gen+Repair)}---T0 plus the execution-guided repair loop that consumes fuzzer/SUT outcomes and emits minimal patches for re-synthesis.
We reuse the RQ1 specification metrics (precision/recall by element) the RQ2 execution metrics (MA, TA, RtC), reporting only deltas (T1--T0). Beyond RQ1/2, we summarize repair dynamics with two macro plots: (i) a survival curve showing the fraction of grammars not yet fixed versus repair round (0--7), aggregated over all five protocols (grammars unfixed after round~7 are treated as censored) (\Cref{fig:rq3-survival}); and (ii) stacked bars of fixes per round (1--7), stratified by the dominant prior failure class (\emph{Grammar Syntax Error}, \emph{Trace Mismatch}, and \emph{Grammar Coverage Miss}) to reveal which rounds contribute most and which errors are most amenable to repair (\Cref{fig:rq3-bar}).

In \Cref{fig:rq3-bar}, each bar partitions the grammars at risk at the start of round r into four disjoint outcomes in that round. Bars sum to 100\% per round.
This answers exactly “in round r, how many got fixed and how many fell into each error category,”
Interpretation. The loop is most effective early, converting many Grammar Syntax Error to successes by round 2–3. What remains are semantic/behavioral mismatches that are harder to resolve automatically. This suggests prioritizing syntax-aware repairs in early rounds and introducing trace-semantics aids (e.g., exemplar alignment, stricter oracle checks, or targeted prompts) for later rounds.

 Most errors are Grammar Syntax Error and fixable quickly; the hard tail is T-Trace Mis-match. The figure quantifies both the when (early vs late) and the what (error class) of repair effectiveness, explaining the survival curve’s early drop and late flattening.

\conclusion{The \AUTOSPEC repair loop significantly improves specification quality, fixing syntax as well as semantic errors.}

\newcommand{\dpp}[1]{\,{\scriptsize($\Delta$\,#1)}} 
\begingroup
\colorlet{Goldenrod}{RoyalBlue!20!white}
\begin{table}[t]
  \footnotesize
  \renewcommand{\arraystretch}{1.1}
  \setlength{\tabcolsep}{4pt}
  \caption{RQ3: Recall of I/O Grammar Extraction} 
  \label{tab:rq3_recall}

  \rowcolors{2}{gray!20}{white}
  \begin{tabularx}{\linewidth}{@{}
      l
      >{\centering\arraybackslash}X
      >{\centering\arraybackslash}X
    @{}}

    \rowcolor{Goldenrod}
    \multicolumn{1}{@{}c}{\textbf{Protocol}} &
    \textbf{Client Msg Type} &
    \textbf{Server Msg Type} \\

POP3        & 12/14 (85.7\%)\dpp{-14.3\%} & 6/9 (66.7\%)\dpp{-33.3\%} \\
SMTP        & 11/11 (100.0\%)\dpp{0}      & 5/11 (45.5\%)\dpp{-18.2\%} \\
IMAP        & 25/30 (83.3\%)\dpp{0}       & 10/17 (58.8\%)\dpp{-11.8\%} \\
FTP         & 25/31 (80.6\%)\dpp{0}       & 4/15 (26.7\%)\dpp{-40.0\%} \\
ManageSieve & 14/14 (100.0\%)\dpp{0}      & 6/6 (100.0\%)\dpp{0} \\

  \end{tabularx}

  \vspace{0.25em}
  \captionsetup{font=footnotesize}
\end{table}
\begingroup
\colorlet{Goldenrod}{RoyalBlue!20!white}
\begin{table}[t]
  \footnotesize
  \renewcommand{\arraystretch}{1.1}
  \setlength{\tabcolsep}{4pt}
  \caption{RQ3: Precision of I/O Grammar Extraction}
  \label{tab:rq3_precision}

  \rowcolors{2}{gray!20}{white}
  \begin{tabularx}{\linewidth}{@{}
      l
      >{\centering\arraybackslash}X
      >{\centering\arraybackslash}X
    @{}}

    \rowcolor{Goldenrod}
    \multicolumn{1}{@{}c}{\textbf{Protocol}} &
    \textbf{Client Msg Type} &
    \textbf{Server Msg Type} \\

POP3        & 12/12 (100.0\%)\dpp{0} & 6/6 (100.0\%)\dpp{0} \\
SMTP        & 11/11 (100.0\%)\dpp{0} & 5/5 (100.0\%)\dpp{0} \\
IMAP        & 25/25 (100.0\%)\dpp{0} & 10/10 (100.0\%)\dpp{0} \\
FTP         & 25/26 (96.2\%)\dpp{0}  & 4/7 (57.1\%)\dpp{-14.3\%} \\
ManageSieve & 14/14 (100.0\%)\dpp{0} & 6/6 (100.0\%)\dpp{0} \\

  \end{tabularx}

  \vspace{0.25em}
  \captionsetup{font=footnotesize}
\end{table}

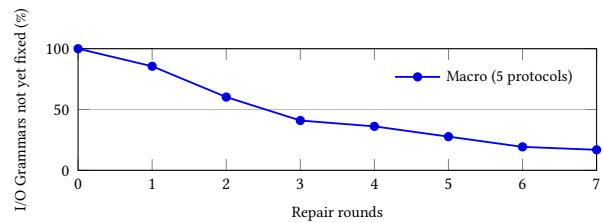
\begin{figure}[t]
  \centering
  \begin{tikzpicture}
    \begin{axis}[
      width=\linewidth,
      height=3.2cm,                    
      xmin=0, xmax=7,
      ymin=0, ymax=100,
      xtick={0,...,7},
      ytick={0,50,100},                
      xlabel={Repair rounds},
      ylabel={I/O Grammars not yet fixed (\%)},
      grid=major,                      
      ymajorgrids=true,
      xmajorgrids=false,
      legend style={
        at={(0.98,0.92)},              
        anchor=north east,
        draw=none, fill=none,
        font=\scriptsize,
        /tikz/every even column/.style={column sep=0.5em}
      },
      legend cell align=left,
      tick label style={font=\scriptsize,/pgf/number format/fixed},
      label style={font=\scriptsize},
    ]
      
      \addplot+[mark=*, mark size=1.4pt, line width=0.7pt] 
        coordinates {(0,100.0) (1,85.54) (2,60.24) (3,40.96) (4,36.14) (5,27.71) (6,19.28) (7,16.87)};
      \addlegendentry{Macro (5 protocols)}
    \end{axis}
  \end{tikzpicture}
  \vspace{-1mm}
  \caption{RQ3: Survival of unfixed grammars}
  \label{fig:rq3-survival}
  \vspace{-2mm}
\end{figure}

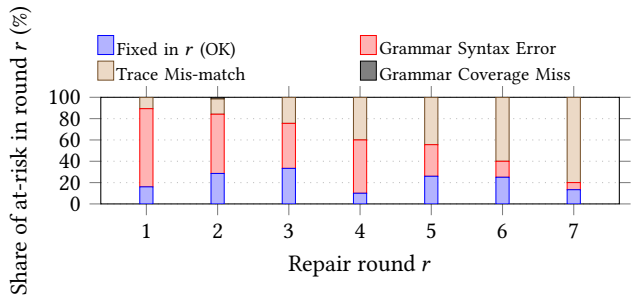
\begin{figure}[t]
  \centering
  \begin{tikzpicture}
    \begin{axis}[
      width=\linewidth, height=3.0cm,        
      ybar stacked, bar width=5pt,            
      xmin=0.5, xmax=7.5,
      ymin=0, ymax=100,
      xtick={1,...,7},
      xlabel={Repair round $r$},
      ylabel={Share of at-risk in round $r$ (\%)},
      ymajorgrids,                           
      grid style={dotted},                    
      tick align=outside,
      tick style={thin},
      legend style={
        at={(0.5,1.05)},anchor=south,draw=none,fill=none,
        legend columns=2,
        nodes={align=left, text width=3.2cm, inner sep=1pt},
        font=\footnotesize
      },
      tick label style={/pgf/number format/fixed},
      every axis plot/.append style={line width=0.3pt},  
      enlarge x limits=0.02                             
    ]

      \addplot+ coordinates {(1,16.0) (2,28.57) (3,33.33) (4,10.0) (5,25.93) (6,25.0) (7,13.33)}; 
      \addlegendentry{Fixed in $r$ (OK)}

      \addplot+ coordinates {(1,73.33) (2,55.56) (3,42.22) (4,50.0) (5,29.63) (6,15.0) (7,6.67)}; 
      \addlegendentry{Grammar Syntax Error}

      \addplot+ coordinates {(1,10.67) (2,14.29) (3,24.44) (4,40.0) (5,44.44) (6,60.0) (7,80.0)}; 
      \addlegendentry{Trace Mis-match}

      \addplot+ coordinates {(1,0.0) (2,1.59) (3,0.0) (4,0.0) (5,0.0) (6,0.0) (7,0.0)}; 
      \addlegendentry{Grammar Coverage Miss}

    \end{axis}
  \end{tikzpicture}
  \vspace{-2mm}
  \caption{Per-round outcomes among grammars at risk at the start of round $r$. Bars split into ``fixed in $r$'' vs failure classes.}
  \label{fig:rq3-bar}
  \vspace{-2mm}
\end{figure}

\subsection{RQ4: Naïve prompting}
\emph{Can an LLM, without RFC input, produce useful I/O grammars?}
We removed our entire RFC retrieval pipeline and evaluated three "no-RFC"  settings under identical time and token budgets: (a) a zero-shot prompt providing only the target grammar schema, (b) a few-shot prompt that augmented the schema with two miniature, out-of-domain grammar examples, and (c) a prompt supplying only a bare list of command names.
The results from this experiment were unambiguous, as summarized in Table~\ref{tab:rq4_results}. 
All three approaches failed to produce a valid, executable I/O grammar. The zero-shot approach yielded no usable artifacts, failing even to generate syntactically correct (In the table, \# of valid output/ \# of trials). 
While the few-shot prompts produced structurally sound grammars, their content was hallucinated, bearing no resemblance to the actual  POP3 protocol. The command-list prompt leveraged the model's superficial pre-trained knowledge to correctly model the initial \texttt{USER/PASS} authentication sequence. 

None of the generated grammars could be used for actual testing. 
This outcome confirms our central thesis: an LLM's generalized, pre-trained knowledge is fundamentally insufficient for this task. 
The detailed, normative context from a source like an RFC is not merely helpful, but indispensable.
\conclusion{Naïve prompting without RFC input or repair fails to produce useful protocol specifications (I/O grammars).}

\begingroup
\colorlet{Goldenrod}{RoyalBlue!20!white}
\begin{table}[t]
  \footnotesize
  \renewcommand{\arraystretch}{1.1}
  \setlength{\tabcolsep}{4pt}
  \caption{RQ4: POP3 I/O Grammar Generation without RFCs}
  \label{tab:rq4_results}

  \rowcolors{2}{gray!20}{white}
  \begin{tabularx}{\linewidth}{@{}
      l
      >{\centering\arraybackslash}X
      >{\centering\arraybackslash}X
      >{\centering\arraybackslash}X
    @{}}  

    \rowcolor{Goldenrod}
    \multicolumn{1}{@{}c}{\textbf{Metric}} &
    \textbf{Zero-shot} &
    \textbf{Few-shot} &
    \textbf{Command-list} \\

    Syntactically valid & 0/5 & 2/5 & 3/5 \\
    Executable          & No  & No  & No  \\
    Command coverage    & 0\% & \textless 10\% (generic) & \textasciitilde 40\% \\
    State transitions   & None & Hallucinated & Incorrect \\
    Semantic constraints& None & None & None \\
  \end{tabularx}

  \vspace{0.25em}
  \captionsetup{font=footnotesize}
\end{table}
\endgroup

\subsection{Threats and Limitations}

\iflong
\else
Our study faces the following threats and limitations:
\fi

\subsubsection{Retrieval Quality}

RFCs are structurally diverse: a single normative rule can be split across 
\iflong
\else
several sections.
\fi
We use an off-the-shelf BM25 retriever. We did not fine-tune a retriever or build an RFC-specific terminology base due to the expert effort and lack of labeled data. As a result, important clauses may be missed or misranked, and the synthesized I/O grammars may omit details, reducing precision and completeness. To mitigate this, we will (1) compare alternative retrieval methods (hybrid sparse+dense, multi-stage reranking), (2) explore RFC-specific indexing strategies (section-aware fields, cross-reference expansion), and (3) use our growing RFC clause-I/O grammar pairs to fine-tune models via supervised and active learning.

\subsubsection{Repair Overfitting to Implementations}

The execution-guided repair loop updates the specification based on SUT responses. This can drift toward an implementation-specific formal specification that validates the tested SUT rather than the RFC. We mitigate this by keeping clause-level trace links and validating each repair against the source RFC text. Residual overfitting may remain, especially when multiple implementations share the same deviation.

\subsubsection{Prior Knowledge and Contamination}

LLMs may have \emph{prior knowledge} of popular protocols; this can include memorized protocol specifications or public implementations.
However, as shown in RQ4, prior protocol knowledge is not sufficient to obtain comprehensive formal protocol specifications: naïve one-shot prompting without RFC context does not yield faithful or complete I/O grammars, underscoring the necessity of our structured pipeline.
\iflong
\else
We
\fi
do not see such contamination as harmful; on the contrary, it can help to improve initial synthesis quality and to bootstrap a corpus of natural-language-to-formal-grammar pairs, which in the long run will allow us to learn models for less common protocols.

\subsubsection{Protocol Scope and Generalizability}

Our evaluation covers five ASCII-oriented session protocols (SMTP, POP3, IMAP, FTP, ManageSieve). Extending to binary, encrypted, or real-time stacks (e.g., TLS 1.3, QUIC, SOME/IP) adds challenges: bit/field framing, cryptographic handshakes, and timing-coupled state. We are actively expanding the pipeline to these families; results will follow.

\section{Conclusion and Future Work}
\label{sec:conclusion}

Software specifications written in natural language are ambundant, but need to be \emph{formalized} to be used for automated test generation---a labor-intensive and error-prone process.
We present \AUTOSPEC, an LLM-agentic pipeline that automatically translates natural-language RFC protocol specifications into executable formal specifications with sentence-level traceability.
The resulting I/O grammars can be directly used for effective session-level protocol testing with high precision and diversity.

For us at Volkswagen, this is an important first step towards widespread automated test generation using precisely defined interactions and oracles.
Our future work will focus on these topics:

\begin{description}
  \item[Building a corpus of RFC clauses and I/O grammar pairs.] We want to create a significant corpus of RFC clauses and corresponding I/O grammar fragments to fine-tune retrieval and generation models.
  \item[Complex protocol stacks.] Complex protocols such as TLS~1.3, QUIC, or SOME/IP bring new challenges such as bit/field framing, crypto handshakes, and timing-coupled state.
  \item[Bootstrapping translation models.] Leveraging the growing corpus, we will fine-tune translation models combining supervised and unsupervised learning techniques, improving translation accuracy and reducing the need for repair.
  \item[Ad-hoc specifications.] Once our models have gained sufficient understanding of how natural language translates into formal models and constraints, we can have programmers express  requirements in natural language
\iflong
\else
  too:
\fi
  ``Now test this component using a voltage of 2mV.''
\end{description}

The experimental data is available as an electronic appendix to this paper;
we plan to make \AUTOSPEC publicly available as well:
\begin{center}
\url{https://drive.google.com/drive/folders/1TFRsOUbNby-tg3SIDORk5tZr21GsFiZV}
\end{center}

\bibliographystyle{ACM-Reference-Format}
\bibliography{references.bib}

\end{document}